\documentclass[12pt]{article}
\usepackage[english]{babel}
\usepackage[latin1]{inputenc}
\usepackage{amsfonts,amssymb,amsmath, epsfig}
\usepackage{color,graphicx,graphics,psfrag}
\usepackage{amsmath,amstext,amssymb,amsfonts, amscd}
\usepackage{hyperref}           

\def\Box{\leavevmode\vbox{\hrule
     \hbox{\vrule\kern4pt\vbox{\kern4pt}%
           \vrule}\hrule}}
\def\blackbox{\leavevmode\vrule height 5pt width 4pt depth 0pt\relax}
\def\endproof{\null\hfill {$\blackbox$}\bigskip}

\newcounter{appendix}
\def\appendix{\advance\c@appendix by 1
   \def\thesection{\Alph{section}}
   \ifnum\c@appendix=1 \setcounter{section}{-1} \fi
   \@startsection {section}{1}{\z@}{-3.5ex plus -1ex minus 
   -.2ex}{2.3ex plus .2ex}{\Large\bf}}

\def\paragraph#1{{\bf #1\ }}

\newtheorem{lemma}{Lemma}[section]  

\newtheorem{theorem}[lemma]{Theorem}

\newtheorem{definition}[lemma]{Definition}

\newtheorem{proposition}[lemma]{Proposition}

\newtheorem{remark}{Remark}[section]

\title{Hydrodynamics of self-alignment interactions with precession and derivation of the Landau-Lifschitz-Gilbert equation} 
\author{Pierre Degond$^{(1,2)}$, Jian-Guo Liu$^{(3)}$} 
\date{} 
\begin{document}

\maketitle

\vspace{0.5 cm}

\begin{center}
1-Université de Toulouse; UPS, INSA, UT1, UTM ;\\ 
Institut de Mathématiques de Toulouse ; \\
F-31062 Toulouse, France. \\
2-CNRS; Institut de Mathématiques de Toulouse UMR 5219 ;\\ 
F-31062 Toulouse, France.\\
email: pierre.degond@math.univ-toulouse.fr; 
\end{center}

\begin{center}
3- Department of Physics and Department of Mathematics\\
Duke University\\
Durham, NC 27708, USA\\
email: jliu@phy.duke.edu
\end{center}

\vspace{0.5 cm}
\begin{abstract}
We consider a kinetic model of self-propelled particles with alignment interaction and with precession about the alignment direction. We derive a hydrodynamic system for the local density and velocity orientation of the particles. The system consists of the conservative equation for the local density and a non-conservative equation for the orientation. First, we assume that the alignment interaction is purely local and derive a first order system. However, we show that this system may lose its hyperbolicity. Under the assumption of weakly non-local interaction, we derive diffusive corrections to the first order system which lead to the combination of a heat flow of the harmonic map and Landau-Lifschitz-Gilbert dynamics. In the particular case of zero self-propelling speed, the resulting model reduces to the phenomenological Landau-Lifschitz-Gilbert equations. Therefore the present theory provides a kinetic formulation of classical micromagnetization models and spin dynamics. 
\end{abstract}

\medskip
\noindent
{\bf Acknowledgements:} The authors wish to acknowledge the hospitality of Mathematical Sciences Center and Mathematics Department of Tsinghua University where this research was performed. The research of J.-G. L. was partially supported by NSF grant DMS 10-11738.


\vskip 0.4cm

\setcounter{equation}{0}
\section{Introduction}
\label{sec_intro}

The setting of this paper is the same as in \cite{DM} and considers a kinetic model of self-propelled particles in three dimensions and its hydrodynamic limit. The particles are supposed to move with the same constant speed and their velocity orientation (which are vectors of the two-dimensional sphere ${\mathbb S}^2$) tend to align to the local average orientation, with some noise. This model has been proposed as a kinetic formulation of the Vicsek particle model \cite{Vicsek}. In this paper, we investigate how the addition of a precession force around the local average orientation modifies the resulting kinetic and hydrodynamic models. 

The derivation of the hydrodynamic model proceeds like in \cite{DM} and relies on both the determination of local equilibria in the form of Von-Mises-Fischer (VMF) distributions and on the new concept of 'Generalized Collision Invariants' (GCI) which allows for a non-conservative closure. In the present paper, we show that the VMF equilibria are still equilibrium states of the system with precession and we extend the derivation of the GCI's to this system. As in \cite{DM}, the resulting hydrodynamics (later on referred to as hydrodynamic model for self-alignment with precession) consists of a conservation law for the density and a non-conservative equation for the velocity orientation. By contrast to \cite{DM}, the velocity orientation equation contains an additional term which accounts for the influence of the precession force and which results in the loss of hyperbolicity of the model as soon as the precession force is non-zero. This loss of hyperbolicity occurs when waves propagate in the same direction as that of the average velocity direction. 

In order to stabilize the model, we introduce a weak non-locality in the alignment interaction: the local average which determines the alignment direction is now taken over a certain spatial extension around the particle. If the spatial extension of this local average is scaled like the square root of the alignment mean free path $\varepsilon$, then, the alignment direction possesses a small correction  of order $\varepsilon$ to the local average direction. This correction involves the Laplacian of the velocity orientation and results in diffusion terms in the macroscopic equation. Due to the precession force, this diffusion term has two contribution: the first one corresponds to the heat flow for harmonic maps ; the second one is in the form of a Landau-Lifschitz-Gilbert term. In the particular case of zero speed, the resulting model reduces to the phenomenological Landau-Lifschitz-Gilbert equations. Therefore the present theory provides a kinetic formulation of classical micromagnetization models and spin dynamics \cite{Brown}. 

In the non-local alignment interaction, we also include a small (of order $\varepsilon$) repulsive force which, in the macroscopic model, gives riseto both an additional pressure force and a pressure precession term. 

The Vicsek model \cite{Vicsek} has been proposed as a model for the interaction of individuals among animal societies such as fish schools, bird flocks, herds of mammalians, etc (see also \cite{Aldana_Huepe, Aoki, Couzin, Gregoire_Chate}). It is a particle model (or 'Individual-Based Model' or 'Agent-Based model') which consists in an Ordinary Differential Equation system for the particle positions and velocities. As the number of particles increases, its computational cost increases more than linearly and becomes prohibitive for large particle numbers. The Vicsek model is set up in a time-discrete framework. A time-continuous version of the Vicsek model has been proposed in \cite{DM} and a kinetic model has been deduced from it on formal grounds.  The rigorous derivation of this kinetic model has been performed in \cite{BCC}. In the present paper, we will directly start from the kinetic level. 

For the modeling of systems consisting of a large number of individuals, it is more efficient to use  hydrodynamic models, which describe the system by macroscopic averages. Many such  models exist \cite{Chuang, Dorsogna, Mogilner1, Mogilner2, TB1, TB2} but few of them are rigorously derived from particle or kinetic dynamics. The first rigorous derivation of the hydrodynamic limit of the Vicsek model is done in \cite{DM} (earlier phenomenological derivations can be found in \cite{KRZB,RBKZ,RKZB} but do not lead to the same model). The resulting hydrodynamic model has also been found in the hydrodynamics of the so-called Persistent Turning Walker model \cite{DM2} of fish behavior, where particles interact through curvature control. Diffusive corrections to the hydrodynamic model have also been computed  in \cite{DY} through a Chapman-Enskog expansion up to the second order. Other macroscopic models of swarming particle systems derived from kinetic theory can be found e.g. in \cite{CDP, CKMT}. In particular, variants of the kinetic model of \cite{DM} and its hydrodynamics have been proposed. In \cite{Frouvelle}, the influence of a vision angle and dependency of the alignement frequency upon the local density has been investigated.  In \cite{FL, DFL}, the modification of the normalization constant of the alignment force results in phase transitions from disordered to ordered equilibria as the density increases and reaches a threshold. In the spatially homogeneous case, these phase transition models are analogs of the Doi-Onsager \cite{DE, Onsager} and Maier-Saupe \cite{MS} models for phase transition in polymers. This makes a strong difference in the resulting macroscopic models as compared with the ones we are presenting here. 

The organization of the paper is as follows. In section \ref{sec:setting}, we set up the kinetic model and discuss the equilibria and GCI's. In section \ref{sec:HL}, the hydrodynamic limit is established. The hyperbolicity of the resulting model is investigated in section \ref{sec:hyperb}. Then, the weakly non-local model is derived in section \ref{sec:NL}. Finally, a conclusion is drawn at section \ref{sec:conclu}.

\setcounter{equation}{0}
\section{Setting of the problem and preliminaries}
\label{sec:setting}

\subsection{The kinetic model}
\label{subsec:kinetic}

The starting point is the following kinetic model for a distribution function $f(x,v,t)$, where $x \in {\mathbb R}^3$, $v \in {\mathbb S}^2$, $t \geq 0$: 
\begin{eqnarray}
f^\varepsilon_t + c v \cdot \nabla_x f^\varepsilon &=& \frac{1}{\varepsilon} \nabla_v \cdot \left[ - ( P_{v^\bot} \Omega_{f^\varepsilon} ) f^\varepsilon + d \nabla_v f^\varepsilon + \alpha (\Omega_{f^\varepsilon} \times v) f^\varepsilon \right]  \label{eq:kinetic_1} \\
& : = & \frac{1}{\varepsilon} Q(f^\varepsilon)  \nonumber \\
& : = & \frac{1}{\varepsilon} (Q_0(f^\varepsilon) + R(f^\varepsilon)) , \nonumber 
\end{eqnarray}
where 
\begin{eqnarray*}
Q_0(f) &=&  \nabla_v \cdot \left[ - ( P_{v^\bot} \Omega_{f} ) f + d \nabla_v f  \right],  \\
R(f)   &=& \alpha \nabla_v \cdot \left[  (\Omega_{f} \times v) f \right] .
\end{eqnarray*}
$Q_0(f)$ is the operator studied in \cite{DM}. $R(f)$ is the precession operator about an axis which is that of the average particle velocity. $c \geq 0$ is the speed of the self-propelled particles.  We let $P_{v^\bot} = {\mbox Id} - v \otimes v$ be the projection operator onto the plane normal to $v$ and
\begin{eqnarray*}
& & \hspace{-1cm} \rho_f = \int_{{\mathbb S}^2} f(v) \, dv, \quad \Omega_f = \frac{\int_{{\mathbb S}^2} f(v) \, v \, dv}{|\int_{{\mathbb S}^2} f(v) \, v \, dv|} ,
\end{eqnarray*}
the density and mean velocity direction associated to $f$.
We define the equilibria (Von-Mises-Fischer distribution): 
\begin{eqnarray*}
F_\Omega &=& \frac{\exp (\beta v \cdot \Omega) }{\int_{{\mathbb S}^2} \exp (\beta v \cdot \Omega) \, dv} ,
\end{eqnarray*}
with $\beta = 1/d$. We note the compatibility relation: 
$$ \rho_{\rho F_\Omega} = \rho , \quad \Omega_{\rho F_\Omega} = \Omega. $$
We define
$$ M_f = \rho_f F_{\Omega_f}, $$
the equilibrium distribution associated to $f$.

\subsection{Equilibria}
\label{subsec:equilibria}

In this section, we investigate the solutions of the equation $Q(f)=0$, which are the so-called local equilibria of the kinetic model. 

\begin{proposition}
We have:
$$ Q(f) = 0 \Longleftrightarrow \exists \rho \geq 0, \, \exists \Omega \in {\mathbb S}^2, \quad \mbox{ such that } \quad f = \rho F_\Omega. $$
\label{prop:equilibria}
\end{proposition}

\medskip
\noindent
{\bf Proof:} We note the following properties: 
\begin{eqnarray*}
&& \hspace{-1cm} \int_{{\mathbb S}^2} Q_0(f) \frac{f}{M_f} \, dv = - d \, \int_{{\mathbb S}^2} \left| \nabla_v \left( \frac{f}{M_f} \right) \right|^2 \, M_f \, dv \leq 0,  \\
&& \hspace{-1cm} \int_{{\mathbb S}^2} R(f) \frac{f}{M_f} \, dv = -  \alpha \int_{{\mathbb S}^2} (\Omega_f \times v) \cdot \left[ \frac{\nabla_v f}{M_f} - (P_{v^\bot} \Omega_f) \frac{f}{M_f} \right] \, f(v) \, dv = 0. 
\end{eqnarray*}
The first line is proved in \cite{DM}. To prove the second line, we remark that $\nabla_v M_f = \beta P_{v^\bot} \omega \, M_f$, $(\Omega_f \times v) \cdot (P_{v^\bot} \Omega_f) = 0$ and that the first term can be written as 
\begin{eqnarray*}
&& \hspace{-1cm} - \frac{\alpha^2}{2} \int_{{\mathbb S}^2} \nabla_v  \left[ (\Omega_f \times v) \frac{f^2}{M_f} \right] \, dv = 0. 
\end{eqnarray*}
Then, if $Q(f) = 0$, we have 
\begin{eqnarray*}
&& \hspace{-1cm} 0 = \int_{{\mathbb S}^2} Q(f) \frac{f}{M_f} \, dv = - d \,  \int_{{\mathbb S}^2} \left| \nabla_v \left( \frac{f}{M_f} \right) \right|^2 \, M_f \, dv .
\end{eqnarray*}
This implies that $\frac{f}{M_f}$ is a constant, and that therefore, $f$ is of the form $\rho F_\Omega$ for a given $\rho \geq 0$, $\Omega \in {\mathbb S}^2$. Conversely, by the compatibility relation, it is obvious that $f = \rho F_\Omega$ satisfies $Q(f)=0$. 
\endproof

\begin{remark}
We comment on the connection between this model and rod like polymers. In the case of rod-like polymers, the interaction operator is written 
$$ \nabla_v \cdot [  ( -  \nabla_v \phi + w \times v ) f + d \nabla_v f ] , $$
where $\phi(v)$ is a potential and $w$ is the precession direction. In general, finding equilibria for polymer models is difficult \cite{OT}. Here, we deal with the special case where $\phi (v ) = - v \cdot w$ which makes this computation possible. In general, the computation is possible if the level curves of $\phi$ are invariant under precession rotation. 
\end{remark}

\subsection{Generalized Collisions Invariants}
\label{subsec:GCI}

We recall the definition of a Generalized Collision invariant (GCI) (see \cite{DM}). First, we define the collision operator 
\begin{eqnarray*}
{\mathbb Q}(f,\Omega) =  \nabla_v \cdot \left[ - ( P_{v^\bot} \Omega ) f + d \nabla_v f + \alpha (\Omega \times v) f \right],   
\end{eqnarray*}
where now $\Omega  \in {\mathbb S}^2$ is not necessarily equal to $\Omega_f$. For given $\Omega$, ${\mathbb Q}(\cdot,\Omega)$ is a linear operator and we will also consider its formal adjoint ${\mathbb Q}^*(\cdot,\Omega)$.
Then, we have the 

\begin{definition}
For given $\Omega  \in {\mathbb S}^2$, a function $\psi_\Omega(v)$ is said to be a GCI associated to $\Omega$ iff the following property holds:
$$ \int_{{\mathbb S}^2} {\mathbb Q}(f,\Omega) \, \psi_\Omega \, dv = 0, \quad \quad \forall f \quad \mbox{ such that } \quad \Omega_f = \pm \Omega. $$
\label{def:GCI}
\end{definition}

We note that the condition $\Omega_f = \pm \Omega$ is a linear condition on $f$ which can equivalently be written $A \cdot \int_{{\mathbb S}^2} f(v) \, v \, dv = 0$, for all vectors $A$ such that $A \cdot \Omega = 0$. On the other hand, we have 
$$ \int_{{\mathbb S}^2} {\mathbb Q}(f,\Omega) \, \psi_\Omega \, dv = \int_{{\mathbb S}^2} {\mathbb Q}^*(\psi_\Omega,\Omega) \, f \, dv . $$ 
Therefore, that $\psi_\Omega$ is a GCI is equivalent to 
$$ \exists A \in {\mathbb R}^3 \quad \mbox{such that } \quad A \cdot \Omega = 0 \quad \mbox{and } \quad
{\mathbb Q}^*(\psi_\Omega,\Omega) = A \cdot v, $$
or 
$$  ( P_{v^\bot} \Omega ) \cdot \nabla_v \psi_\Omega + d \Delta_v \psi_\Omega - \alpha (\Omega \times v)  \cdot \nabla_v \psi_\Omega = A \cdot v . $$
In spherical coordinates (with polar axis $\Omega$, latitude $\theta \in [0,\pi]$ and longitude $\varphi \in [0,2\pi]$), this equation is written (dropping the index $\Omega$): 
\begin{eqnarray}
& & \hspace{-1cm} - \sin \theta \, \partial_\theta \psi + d \left[ \frac{1}{\sin \theta} \partial_\theta ( \sin \theta \, \partial_\theta \psi ) + \frac{1}{\sin^2 \theta} \partial_\varphi^2 \psi \right] - \alpha \sin \theta \, \partial_\varphi \psi = \nonumber \\
& & \hspace{6cm} = A_1 \sin \theta \, \cos \varphi + A_2 \sin \theta \, \sin \varphi, \label{eq:elliptic}
\end{eqnarray}
where $A = (A_1,A_2,0)$ are the coordinates of $A$, in a ortho-normal coordinate basis $(e_1,e_2,\Omega)$. 

\begin{proposition}
The set ${\mathcal C}_\Omega$ of the GCI's associated to $\Omega$ is a vector space of dimension $3$ spanned by $\{1, \psi^{(1)}, \psi^{(2)} \}$, where $\psi^{(k)}$ is associated to $A = e_k$, for $k = 1, 2$. Furthermore, 
$$  \psi^{(k)} (\theta, \varphi) = \psi^{(k)}_1 (\theta) \cos \varphi + \psi^{(k)}_2 (\theta) \sin \varphi, \quad k=1,2, $$
and 
$$ \psi^{(2)}_1 = - \psi^{(1)}_2, \quad \psi^{(2)}_2 = \psi^{(1)}_1 . $$
Finally, $(\psi^{(1)}_1 , \psi^{(1)}_2)$ satisfies the following elliptic system: 
\begin{eqnarray}
& & \hspace{-1cm} - \sin \theta \, \partial_\theta \psi^{(1)}_1 + d \left[ \frac{1}{\sin \theta} \partial_\theta ( \sin \theta \, \partial_\theta \psi^{(1)}_1 ) - \frac{1}{\sin^2 \theta} \psi^{(1)}_1 \right] - \alpha \sin \theta \, \psi^{(1)}_2 = \sin \theta, \label{eq:elliptic_1}\\ 
& & \hspace{-1cm} - \sin \theta \, \partial_\theta \psi^{(1)}_2 + d \left[ \frac{1}{\sin \theta} \partial_\theta ( \sin \theta \, \partial_\theta \psi^{(1)}_2 ) - \frac{1}{\sin^2 \theta} \psi^{(1)}_2 \right] + \alpha \sin \theta \, \psi^{(1)}_1 = 0. \label{eq:elliptic_2}
\end{eqnarray}
$(\psi^{(2)}_1 , \psi^{(2)}_2)$ satisfies the same system with right hand sides $0$ and $\sin \theta$ for the first and second equation respectively. The elliptic system has a unique zero-average solution, which defines $\psi^{(k)}$ uniquely. 
\label{prop:GCI}
\end{proposition}

\noindent
{\bf Proof:} The reduction of (\ref{eq:elliptic}) to (\ref{eq:elliptic_1}), (\ref{eq:elliptic_2}) follows from using Fourier transform in the $\varphi$ variable. To solve this system, we construct the complex-valued function $\tilde \psi^{(k)} (\theta) = \psi^{(k)}_1 (\theta) + i \psi^{(k)}_2 (\theta)$. $\tilde \psi^{(1)}$ satisfies the complex-valued elliptic problem: 
\begin{eqnarray*}
& & \hspace{-1cm} - \sin \theta \, \partial_\theta \tilde \psi^{(1)} + d \left[ \frac{1}{\sin \theta} \partial_\theta ( \sin \theta \, \partial_\theta \tilde \psi^{(1)} ) - \frac{1}{\sin^2 \theta} \tilde \psi^{(1)} \right] + i  \alpha \sin \theta \, \tilde \psi^{(1)} = \sin \theta, 
\end{eqnarray*}
whereas $\tilde \psi^{(2)}$ satisfies the same equation with right-hand side $ i \sin \theta$. Since the imaginary part of the operator is skew-adjoint, it does not modify the energy identity. Therefore, the same theory as in \cite{DM} applies and shows the existence and uniqueness of the solution in the space of zero average functions. The remaining part of the statement follows easily. 
\endproof

\setcounter{equation}{0}
\section{Hydrodynamic limit}
\label{sec:HL}

The following theorem establishes the limit $\varepsilon \to 0$. 

\begin{theorem}
We suppose that $f^\varepsilon \to f^0$ as smoothly as needed. Then, we have 
\begin{equation}
f^0 = \rho F_\Omega,  
\label{eq:f0}
\end{equation}
where $\rho = \rho(x,t)$ and $\Omega = \Omega(x,t)$ satisfy the following system 
\begin{eqnarray}
&& \hspace{-1cm} \partial_t \rho + c c_1 \nabla_x (\rho \Omega) = 0, \label{eq:rho_limit} \\
&& \hspace{-1cm} \rho (\partial_t \Omega + c c_2 \,  \cos \delta \,  (\Omega \cdot \nabla_x) \Omega + c c_2 \, \sin \delta \, \Omega \times ((\Omega \cdot \nabla_x) \Omega) ) + c d \, P_{\Omega^\bot} \nabla_x \rho = 0 . \label{eq:omega_limit}
\end{eqnarray}
The coefficient $c_1$ is defined by 
\begin{eqnarray}
&& \hspace{-1cm} c_1 = \int_{{\mathbb S}^2} F_\Omega(v) \, (v \cdot \Omega) \, dv  = \frac{\int_{{\mathbb S}^2} \exp (\beta v \cdot \Omega) \, (v \cdot \Omega) \, dv }{\int_{{\mathbb S}^2} \exp (\beta v \cdot \Omega) \, dv} .
\label{eq:coef_c1}
\end{eqnarray}
The coefficients $c_2$ and $\delta$ are defined as follows. First, define $a_k$ and $b_k$ for $k=1,2$ by 
\begin{eqnarray}
&& \hspace{-1cm} a_k = \frac{1}{2} \int_{\theta \in [0,\pi]} F_\Omega(\cos \theta) \, \psi_k^{(1)} (\theta)  \sin^2 \theta \, d \theta, \quad k=1,2 , \label{eq:coef_ak} \\
&& \hspace{-1cm} b_k = \frac{1}{2} \int_{\theta \in [0,\pi]} F_\Omega(\cos \theta) \, \psi_k^{(1)} (\theta) \cos \theta \,  \sin^2 \theta \, d \theta, \quad k=1,2, \label{eq:coef_bk}
\end{eqnarray}
where $\psi_k^{(1)}$ are the GCI's found in proposition \ref{prop:GCI}. Then, introducing $(\rho_a,\theta_a)$, $(\rho_b,\theta_b)$ the polar coordinates of the vectors $(a_1,a_2)$ and $(b_1,b_2)$, i.e. 
\begin{eqnarray*}
&& \hspace{-1cm} a_1 + i a_2 = \rho_a e^{i \theta_a}, \quad b_1 + i b_2 = \rho_b e^{i \theta_b}
, 
\end{eqnarray*}
$c_2$ and $\delta$ are defined by
\begin{eqnarray}
&& \hspace{-1cm} c_2 = \frac{\rho_b}{\rho_a}, \quad \delta = \theta_b - \theta_a. \label{eq:coef_rho_theta}
\end{eqnarray}
\label{thm:limit}
\end{theorem}

\begin{remark}
Model (\ref{eq:rho_limit}), (\ref{eq:omega_limit}) will be referred to as the hydrodynamic model for self-alignment interactions with precession.
\end{remark}

\begin{remark}
We note the following identities: 
\begin{eqnarray*}
&& \hspace{-1cm} (\Omega \cdot \nabla_x) \Omega= (\nabla_x \times \Omega) \times \Omega, \\ 
&& \hspace{-1cm} \Omega \times ( (\nabla_x \times \Omega) \times \Omega ) = P_{\Omega^\bot}  (\nabla_x \times \Omega). 
\end{eqnarray*}
\end{remark}

\begin{remark}
The reason for keeping the dependence upon the speed $c$ explicit is that we will later investigate the case $c=0$. 
\end{remark}

\medskip
\noindent
{\bf Proof.} Letting $\varepsilon \to 0$ in (\ref{eq:kinetic_1}) and using proposition \ref{prop:equilibria}, we get that $Q(f^\varepsilon) \to 0 = Q(f^0)$, and that eq. (\ref{eq:f0}) is satisfied. Now, we look for the equations satisfied by $(\rho, \Omega)$ and for this purpose, we use the GCI.

According to Proposition \ref{prop:GCI}, $1$ is a classical collision invariant. Therefore, multiplying (\ref{eq:kinetic_1}) by $1$ and letting $\varepsilon \to 0$, we get the mass conservation equation (\ref{eq:rho_limit}). 

Now, we successively multiply (\ref{eq:kinetic_1}) by the GCI's $\psi^{(k)}$, $k=1, 2$ associated to the polar vector $\Omega_{f^\varepsilon}$. From the fact that $\psi^{(k)} \in {\mathcal C}_{\Omega_{f^\varepsilon}}$, we get that 
$$ \int_{{\mathbb S}^2} Q(f^\varepsilon) \psi^{(k)} \, dv = \int_{{\mathbb S}^2} {\mathbb Q}(f^\varepsilon, \Omega_{f^\varepsilon}) \psi^{(k)} \, dv = 0 , \quad k=1, 2, $$
thanks to definition \ref{def:GCI}. We deduce that 
$$  \int_{{\mathbb S}^2} T(f^\varepsilon) \psi^{(k)} \, dv = 0, \quad k=1, 2, $$
where $Tf = \partial_t f + c v \cdot \nabla_x f $ is the transport operator. Letting $\varepsilon \to 0$, we deduce that 
\begin{equation} 
{\mathcal T}^{(k)} := \int_{{\mathbb S}^2} T(\rho F_\Omega) \psi^{(k)} \, dv = 0, \quad k=1, 2. 
\label{eq_GCI_1}
\end{equation}
We note that 
\begin{eqnarray}
&& \hspace{-1cm} \psi^{(1)} = \psi_1 (\theta) \cos \varphi + \psi_2 (\theta) \sin \varphi , \label{eq_psik_1} \\
&& \hspace{-1cm} \psi^{(2)} = - \psi_2 (\theta) \cos \varphi + \psi_1 (\theta) \sin \varphi, \label{eq_psik_2}
\end{eqnarray}
where we define $\psi_1 = \psi^{(1)}_1$ and $\psi_2 = \psi^{(1)}_2$ for simplicity.

A simple computation \cite{DM} shows that 
$$ T(\rho F_\Omega) = F_\Omega \left\{ \partial_t \rho + c (v \cdot \nabla_x) \rho + \beta \rho (v \cdot (\partial_t + c v \cdot \nabla_x) \Omega) \right\}. $$
We decompose 
$$ v = v_\bot + v_\parallel, \quad v_\bot = P_{\Omega^\bot} v, \quad v_\parallel = (v \cdot \Omega) \Omega. $$
In spherical coordinate system, we have
$$ v_\bot = (\sin \theta \cos \varphi, \sin \theta \sin \varphi, 0)^T, \quad v_\parallel = (0,0, \cos \theta).$$
Therefore, $v_\bot$ is an odd degree trigonometric polynomial of $\varphi$, while $v_\parallel$ is an even degree such polynomial. We decompose $T(\rho F_\Omega)$ into even and odd degree trigonometric polynomials in $\varphi$ using the above defined decomposition of $v$. We get: 
\begin{eqnarray}
&& \hspace{-1cm} T(\rho F_\Omega) = T_e + T_o, \label{eq_GCI_2}\\
&& \hspace{-1cm}  T_o = F_\Omega \left\{ c v_\bot \cdot \nabla_x \rho + \beta \rho v_\bot \cdot (\partial_t + c (v \cdot \Omega) \Omega \cdot \nabla_x) \Omega \right\}, \nonumber
\end{eqnarray}
where we have used that $v_\parallel \cdot ((v_\bot \cdot \nabla_x) \Omega) = 0$ because $|\Omega|=1$. 
We can write 
$$ T_o =  F_\Omega \, v_\bot \cdot D(v \cdot \Omega) =  F_\Omega \, \sin \theta \, (\cos \varphi \, D_1(\cos \theta) + \sin \varphi \, D_2(\cos \theta)), $$
with 
\begin{equation} 
D(v \cdot \Omega) = c \nabla_x \rho + \beta \rho (\partial_t + c (v \cdot \Omega) \Omega \cdot \nabla_x) \Omega,
\label{eq_GCI_3}
\end{equation}
and $D = (D_1,D_2,D_3)$ are the coordinates of $D$ in the cartesian coordinate basis associated to the definitions of the angles $(\theta, \varphi)$. 

Now, inserting (\ref{eq_GCI_2}), (\ref{eq_GCI_3}) into (\ref{eq_GCI_1}), and noting that the odd degree polynomial in $\varphi$ vanish away upon integration with respect to $\varphi \in [0, 2 \pi]$, we get, 
\begin{eqnarray*}
&& \hspace{-1cm}  {\mathcal T}^{(1)} = \int_{\theta \in [0,\pi], \varphi \in [0, 2 \pi]} F_\Omega(\cos \theta) \, (\cos \varphi \, D_1(\cos \theta) + \sin \varphi \, D_2(\cos \theta)) \\
&& \hspace{6cm}  (\psi_1 (\theta) \cos \varphi + \psi_2 (\theta) \sin \varphi)  \sin^2 \theta \, d \theta \, d \varphi , \\
&& \hspace{-1cm} 
{\mathcal T}^{(2)} = \int_{\theta \in [0,\pi], \varphi \in [0, 2 \pi]} F_\Omega(\cos \theta) \, (\cos \varphi \, D_1(\cos \theta) + \sin \varphi \, D_2(\cos \theta)) \\
&& \hspace{6cm}  (- \psi_2 (\theta) \cos \varphi + \psi_1 (\theta) \sin \varphi)  \sin^2 \theta \, d \theta \, d \varphi .
\end{eqnarray*}
Performing the $\varphi$ integration leads to: 
\begin{eqnarray}
&& \hspace{-1cm}  {\mathcal T}^{(1)} = \frac{1}{2} \int_{\theta \in [0,\pi]} F_\Omega(\cos \theta) \, ( D_1(\cos \theta) \psi_1 (\theta) + D_2(\cos \theta) \psi_2 (\theta) )  \sin^2 \theta \, d \theta  , \label{eq_GCI_4} \\
&& \hspace{-1cm} 
{\mathcal T}^{(2)} = \frac{1}{2} \int_{\theta \in [0,\pi]} F_\Omega(\cos \theta) \, ( - D_1(\cos \theta) \psi_2 (\theta) + D_2(\cos \theta) \psi_1 (\theta) )  \sin^2 \theta \, d \theta . \label{eq_GCI_5}
\end{eqnarray}
Now, introducing the matrices 
\begin{eqnarray*}
&& \hspace{-1cm} [a] = \left( \begin{array}{cc} a_1 & a_2 \\ -a_2 & a_1  \end{array} \right), \quad [b] = \left( \begin{array}{cc} b_1 & b_2  \\ -b_2 & b_1  \end{array} \right), 
\end{eqnarray*}
and 
\begin{eqnarray*}
&& \hspace{-1cm} [A] = \left( \begin{array}{ccc} a_1 & a_2 & 0 \\ -a_2 & a_1 & 0 \\ 0 & 0 & 0  \end{array} \right), \quad [B] = \left( \begin{array}{ccc} b_1 & b_2 & 0 \\ -b_2 & b_1 & 0 \\ 0 & 0 & 0  \end{array} \right), 
\end{eqnarray*}
with $a_k$ and $b_k$ ($k=1,2$) defined at (\ref{eq:coef_ak}), (\ref{eq:coef_bk}) and inserting (\ref{eq_GCI_3}) into (\ref{eq_GCI_4}), (\ref{eq_GCI_5}) leads to 
\begin{eqnarray}
&& \hspace{-1cm} \left( \begin{array}{c} {\mathcal T}^{(1)} \\ {\mathcal T}^{(2)} \\ 0  \end{array} \right) = [A] P_{\Omega^\bot} ( c \nabla_x \rho + \beta \rho \partial_t \Omega) + \beta \rho c [B] P_{\Omega^\bot} ((\Omega \cdot \nabla_x) \Omega) =0 . 
\label{eq:matrix_eq}
\end{eqnarray}
The matrix $[a]$ being a similitude matrix, it is invertible and we can introduce the matrix 
$$ [c_2] = [a]^{-1} \, [b] =  \left( \begin{array}{cc} c_{21} & c_{22} \\ -c_{22} & c_{21}  \end{array} \right), $$
and, 
\begin{eqnarray*}
&& \hspace{-1cm} [C_2] = \left( \begin{array}{ccc} c_{21} & c_{22} & 0 \\ -c_{22} & c_{21} & 0 \\ 0 & 0 & 0  \end{array} \right). 
\end{eqnarray*}
Multiplying the first two lines of the vector equation (\ref{eq:matrix_eq}) by $d [a]^{-1}$, we get:
\begin{eqnarray}
&& \hspace{-1cm} \rho (\partial_t \Omega + c [C_2] (\Omega \cdot \nabla_x) \Omega ) + c d P_{\Omega^\bot} \nabla_x \rho = 0 . \label{eq_GCI_6} 
\end{eqnarray}
This is the momentum equation. 

The equation can be transformed using the fact that $[b]$ and consequently $[c_2]$ are similitude matrices. Therefore, we can write
\begin{equation} 
[c_2] = c_2 R_\delta, 
\label{eq_rotation}
\end{equation}
where $c_2$ and $\delta$ are given by  (\ref{eq:coef_rho_theta}) and
$$ R_\delta = \left( \begin{array}{cc} \cos \delta & \sin \delta \\ - \sin \delta & \cos \delta  \end{array} \right), $$
is the rotation matrix of angle $\delta$. Then, (\ref{eq_GCI_6}) can be equivalently written according to (\ref{eq:omega_limit}), which ends the proof. \endproof

\setcounter{equation}{0}
\section{Hyperbolicity of the hydrodynamic model}
\label{sec:hyperb}

In this section, we investigate the hyperbolicity of the hydrodynamic model for self-alignment interactions with precession. For this purpose, we use the spherical coordinates associated to a fixed cartesian basis. In this basis, denoting by $\theta \in [0,\pi]$ the latitude and $\varphi \in [0,2 \pi]$ the longitude, we have 
$$ \Omega = (\sin \theta \, \cos \varphi, \sin \theta \,  \sin \varphi, \cos \theta)^T, $$
and we let $\Omega_\theta$ and $\Omega_\varphi$ be the derivatives of $\Omega$ with respect to $\theta$ and $\varphi$. We note that 
$$ |\Omega_\theta|=1, \quad |\Omega_\varphi|= \sin \theta. $$
We will use the formulas
\begin{eqnarray*}
& & \hspace{-1cm} \nabla_x \cdot \Omega = \Omega_\theta \cdot \nabla_x \theta + \Omega_\varphi \cdot \nabla_x \varphi, \\
& & \hspace{-1cm} P_{\Omega^\bot} a =  (\Omega_\theta \cdot a) \Omega_\theta + \frac{(\Omega_\varphi \cdot a)}{\sin^2 \theta} \Omega_\varphi , \\
& & \hspace{-1cm} (\Omega \cdot \nabla_x) \Omega = ((\Omega \cdot \nabla_x)\theta) \, \Omega_\theta + ((\Omega \cdot \nabla_x)\varphi) \, \Omega_\varphi, \\
& & \hspace{-1cm} \Omega \times (\Omega \cdot \nabla_x) \Omega = \frac{((\Omega \cdot \nabla_x)\theta)}{\sin \theta} \, \Omega_\varphi - \sin \theta ((\Omega \cdot \nabla_x)\varphi) \, \Omega_\theta, \\
& & \hspace{-1cm} \Omega_t = \Omega_\theta \, \theta_t + \Omega_\varphi \, \varphi_t , 
\end{eqnarray*}
where $a$ is an arbitrary vector. 

As in \cite{DM}, we use the time rescaling $t' = c c_1 t$  and introduce 
$$ a = \frac{c_2}{c_1}, \quad \lambda^2 = \frac{d}{c_1}, \quad \hat \rho = \lambda \, \mbox{ln} \rho.  $$
With this change of variables and unknowns, system (\ref{eq:rho_limit}), (\ref{eq:omega_limit}) is written (dropping the primes on $t$):
\begin{eqnarray*}
&& \hspace{-1cm} \hat \rho_t + \Omega \cdot  \nabla_x \hat \rho + \lambda \nabla_x \cdot \Omega = 0, \\
&& \hspace{-1cm} \Omega_t + a \, ( \cos \delta \,  (\Omega \cdot \nabla_x) \Omega +  \sin \delta \, \Omega \times ((\Omega \cdot \nabla_x) \Omega) )  + \lambda \, P_{\Omega^\bot} \nabla_x \hat \rho = 0 , \end{eqnarray*}
or, 
\begin{eqnarray}
&& \hspace{-1cm} \hat \rho_t + \Omega \cdot  \nabla_x \hat \rho + \lambda  (\Omega_\theta \cdot \nabla_x \theta + \Omega_\varphi \cdot \nabla_x \varphi) = 0, \label{eqnv:rho_limit} \\
&& \hspace{-1cm} \theta_t + a \, ( \cos \delta \,  (\Omega \cdot \nabla_x) \theta -  \sin \delta \, \sin \theta \, (\Omega \cdot \nabla_x) \varphi )  + \lambda \, \Omega_\theta \cdot \nabla_x \hat \rho = 0 , \label{eqnv:theta_limit} \\
&& \hspace{-1cm} \varphi_t + a \, ( \cos \delta \,  (\Omega \cdot \nabla_x) \varphi +  \frac{\sin \delta}{ \sin \theta} \, (\Omega \cdot \nabla_x) \theta )  + \lambda \, \frac{\Omega_\varphi \cdot \nabla_x \hat \rho}{\sin^2 \theta} = 0 . \label{eqnv:varphi_limit} 
\end{eqnarray}
In passing, we note a mistake in formula (4.70) of \cite{DM} where the last term should be divided by $\sin^2 \theta$. This mistake does not affect the eigenvalues of the system which are correct. 

Introducing
$$ U = \left( \begin{array}{c} \hat \rho \\ \theta \\ \varphi \end{array} \right), $$
this system is written 
$$ U_t + A(U) U_x + B(U) U_y + C(U) U_z = 0, $$
in cartesian coordinates $x = (x,y,z)$. Assuming translation invariance along the $z$ direction, the system is reduced to 
$$ U_t + C(U) U_z = 0, $$
with 
$$ C(U) = \left( \begin{array}{ccc} 
\cos \theta & - \lambda \, \sin \theta & 0  \\
- \lambda \, \sin \theta & a \, \cos \delta \, \cos \theta & - a \, \sin \delta \, \sin \theta \, \cos \theta \\
0 & \frac{a \, \sin \delta \, \cos \theta}{\sin \theta} & a \, \cos \delta \, \cos \theta 
\end{array} \right). $$
The system is hyperbolic, if and only if the eigenvalues of $C(U)$ are real for all values of $U$. The characteristic polynomial of $C(U)$ is given by
\begin{eqnarray*} 
& & \hspace{-1cm}P(X) = X^3 - (1+2a \cos \delta) \, \cos \theta \, X^2 + (a \, (a+2 \cos \delta) \, \cos^2 \theta - \lambda^2 \sin^2 \theta) \, X + \\
& & \hspace{6cm} + a \, (\lambda^2 \, \cos \delta \, \sin^2 \theta - a \,  \cos^2 \theta) \, \cos \theta .
\end{eqnarray*}
We note that in the case $\sin \delta = 0$, we recover the results of \cite{DM} where all eigenvalues are real. In the present $\sin \delta \not = 0$ case, the roots of $P(X)$ cannot be analytically computed. However, they are analytically computable in the two cases $\theta = 0$ and $\theta = \pi/2$. 

\medskip
\noindent
{\bf Case $\theta = \pi/2$.} Then, $P(X) = X(X - \lambda)(X + \lambda)$ has real roots $X=0$ and $X = \pm \lambda$. These roots are the same as in the $\delta = 0$ case \cite{DM}. 

\medskip
\noindent
{\bf Case $\theta = 0$.} Then, $P(X) = (X-1)((X - a \, \cos \delta)^2 + a^2 \sin^2 \delta)$ has one real root $X=1$ and two complex conjugate roots $X = a e^{\pm i \delta}$. Therefore, as soon as $\sin \delta \not = 0$, the system loses its hyperbolicity near $\theta = 0$. 

Due to the loss of hyperbolicity of the hydrodynamic model, we look for diffusive corrections by introducing a nonlocal evaluation of the alignment direction. This task is performed in the next section.

\setcounter{equation}{0}
\section{Taking into account non-locality in the interaction}
\label{sec:NL}

Now, we suppose that the $\Omega_f$ vector is replaced by a non-local evaluation denoted by $\tilde \Omega_f^\eta$. So, we introduce the following kinetic model: 
\begin{eqnarray}
&& \hspace{-1cm} f_t + c v \cdot \nabla_x f = \frac{1}{\varepsilon} \nabla_v \cdot \left[ - (P_{v^\bot} \tilde \Omega_f^\eta) f  + d \nabla_v f + \alpha ( \tilde \Omega_f^\eta \times v) f \right], 
\label{eq:k2_1}
\end{eqnarray}
where 
\begin{eqnarray*} 
&& \hspace{-1cm} \tilde \Omega_f^\eta = \frac{j_f^\eta + \eta^2 r_f^\eta}{|j_f^\eta + \eta^2 r_f^\eta|}, \\ 
&& \hspace{-1cm} j_f^\eta =  \int_{(x',v') \in {\mathbb R}^3 \times {\mathbb S}^2} K\left(\frac{|x' - x|}{\eta}\right) v' \, f(x',v',t) \, dv' \, dx' , \\
&& \hspace{-1cm} r_f^\eta =  - \nabla_x \int_{(x',v') \in {\mathbb R}^3 \times {\mathbb S}^2} \Phi\left(\frac{|x' - x|}{\eta}\right) \, f(x',v',t) \, dv' \, dx' , 
\end{eqnarray*}
is a force term. The first term (given by $j_f^\eta$) expresses the alignment interaction (like in the Vicsek dynamics) while the second term (given by $r_f^\eta$) expresses the repulsion interaction. 

We denote:
\begin{eqnarray*}
&& \hspace{-1cm} \int_{x' \in {\mathbb R}^{n}} K(|\xi|)  \, d\xi = k_0 , \\
&& \hspace{-1cm} \frac{1}{2n}  \int_{x' \in {\mathbb R}^{n}} K(|\xi|) \, |\xi|^2 \, d\xi = k ,\\
&& \hspace{-1cm} \int_{x' \in {\mathbb R}^{n}} \Phi (|\xi|)  \, d\xi = \phi . 
\end{eqnarray*}
We can always assume that $k_0=1$. Repulsive interaction here means that we assume $\phi \geq 0$. Defining $u_f$ by 
\begin{eqnarray*}
&& \hspace{-1cm} \rho_f u_f  =  \int_{v' \in {\mathbb S}^2} f(v') \, v' \, dv', 
\end{eqnarray*}
we have the following Taylor expansion of $v_f^\eta$: 
$$  \Omega_f^\eta = \Omega_f + \eta^2 \frac{1}{\rho_f |u_f|} \ell_f + o(\eta^2), \quad \ell_f :=P_{\Omega_f^\bot} ( k \Delta (\rho_f u_f) - \phi \nabla_x \rho_f ) . $$
Inserting this expression into the kinetic equation (\ref{eq:kinetic_1}), we get
\begin{eqnarray*}
&& \hspace{-1cm} f_t + c v \cdot \nabla_x f = \frac{1}{\varepsilon} \nabla_v \cdot \left[  -( P_{v^\bot} \Omega_f ) f + d \nabla_v f + \alpha (\Omega_f \times v) f \right]  \nonumber \\
& & \hspace{1.5cm} + \frac{\eta^2}{\varepsilon} \frac{1}{\rho_f |u_f|} \nabla_v \cdot \left[ - ( P_{v^\bot} \ell_f ) f + \alpha (\ell_f \times v) f \right] + o(\frac{\eta^2}{\varepsilon})  . 
\end{eqnarray*}

We now let $\frac{\eta^2}{\varepsilon} = 1$ and we let $\varepsilon \to 0$. Dropping terms of order $o(1)$ or smaller, this leads to the following problem: 
\begin{eqnarray}
&& \hspace{-1cm} f^\varepsilon_t + c v \cdot \nabla_x f^\varepsilon = \frac{1}{\varepsilon} \nabla_v \cdot \left[  -( P_{v^\bot} \Omega_{f^\varepsilon} ) f^\varepsilon + d \nabla_v f^\varepsilon + \alpha (\Omega_{f^\varepsilon} \times v) f^\varepsilon \right]  \nonumber \\
& & \hspace{1.5cm} +  \frac{1}{\rho_{f^\varepsilon} |u_{f^\varepsilon}|} \nabla_v \cdot \left[ - ( P_{v^\bot} \ell_{f^\varepsilon} ) f^\varepsilon + \alpha (\ell_{f^\varepsilon} \times v) f^\varepsilon \right]   
\label{eq:k2_2} \\
&& \hspace{1.7cm} = \frac{1}{\varepsilon} Q(f^\varepsilon) - L(f^\varepsilon), \nonumber
\end{eqnarray}
with 
$$ Lf = - \frac{1}{\rho_f |u_f|} \nabla_v \cdot \left[ - ( P_{v^\bot} \ell_f ) f + \alpha (\ell_f \times v) f \right]. $$

The following theorem establishes the formal $\varepsilon \to 0$ limit.

\begin{theorem}
We suppose that $f^\varepsilon \to f^0$ as smoothly as needed. Then, we have 
\begin{equation}
f^0 = \rho F_\Omega,  
\label{eq:f0_nl}
\end{equation}
where $\rho = \rho(x,t)$ and $\Omega = \Omega(x,t)$ satisfy the following system:
\begin{eqnarray}
&& \hspace{-1cm} \partial_t \rho + c c_1 \nabla_x (\rho \Omega) = 0, \label{eqnl:rho_limit} \\
&& \hspace{-1cm} \rho \{ \partial_t \Omega + c c_2 \cos \delta  \, (\Omega \cdot \nabla_x) \Omega  + c c_2 \sin \delta  \, \Omega \times ((\Omega \cdot \nabla_x) \Omega ) \}  + c d P_{\Omega^\bot} \nabla_x \rho + \nonumber \\
&& \hspace{2cm} + \frac{1}{c_1} \left\{ - (2d  + c_2 \cos \delta) P_{\Omega^\bot} ( k c_1 \Delta (\rho \Omega) - \phi \nabla_x \rho ) + \right. \nonumber \\
&& \hspace{4cm} + \left. (c_2 \sin \delta - \alpha)  ( \Omega \times ( k c_1 \Delta (\rho \Omega) - \phi \nabla_x \rho )) \right\} = 0 .
 \label{eqnl:omega_limit}
\end{eqnarray}
and the coefficients $c_1$, $c_2$ and $\delta$ are the same as in Theroem \ref{thm:limit}. We restrict ourselves to the case $2d  + c_2 \cos \delta \geq 0$ which is the condition for the system to be stable. 
\label{thm:limit_nonlocal}
\end{theorem}

\begin{remark}
A special case is $c=0$, $\rho =1$, $\phi = 0$. This leads to 
\begin{eqnarray*}
&& \hspace{-1cm} \partial_t \Omega  + k (2d  + c_2 \cos \delta) \, \Omega \times (\Omega \times \Delta  \Omega) + k (c_2 \sin \delta - \alpha)  ( \Omega \times  \Delta  \Omega ) = 0 , 
\end{eqnarray*}
which is the classical Landau-Lifschitz-Gilbert equation, provided that $2d  + c_2 \cos \delta \geq 0$. 
\end{remark}

\medskip
\noindent
{\bf Proof.} Since $Lf$ is a divergence in velocity space, the mass conservation equation is unaffected by this additional term compared to section \ref{sec:HL}. The additional contribution of $Lf$ to the momentum equation is a term of the form: 
\begin{equation} 
{\mathcal L}^{(k)} := \int_{{\mathbb S}^2} L(\rho F_\Omega) \psi^{(k)} \, dv , \quad k=1, 2, 
\label{eq_L_1}
\end{equation}
at the left hand side of (\ref{eq:matrix_eq}). We have
$$ L(\rho F_\Omega) = - \frac{1}{c_1} \nabla_v \cdot \left[ - ( P_{v^\bot} \ell ) F_\Omega + \alpha (\ell \times v) F_\Omega \right], $$
where  
$$ \ell := \ell_{\rho F_\Omega} :=P_{\Omega^\bot} ( k c_1 \Delta (\rho \Omega) - \phi \nabla_x \rho ) . $$
Therefore, using Green's formula, we get: 
\begin{eqnarray*} 
{\mathcal L}^{(k)} &=& \frac{1}{c_1} \int_{{\mathbb S}^2} \left[ - ( P_{v^\bot} \ell )  + \alpha (\ell \times v)  \right] \cdot \nabla_v \psi^{(k)} F_\Omega \, dv  \\
&=& \frac{1}{c_1} \int_{{\mathbb S}^2} \left[ -  \ell + \alpha (\ell \times v) \right] \cdot \nabla_v \psi^{(k)} F_\Omega \, dv = 0 \\
&=& \frac{1}{c_1} \left( \int_{{\mathbb S}^2} \left[ -   \nabla_v \psi^{(k)} + \alpha (v \times \nabla_v \psi^{(k)}) \right] F_\Omega \, dv  \right) \cdot \ell  .
\end{eqnarray*}

Now, we use the formulas: 
\begin{eqnarray*} 
& & \int_{{\mathbb S}^2} \nabla_v g \, dv = 2 \int_{{\mathbb S}^2} v g \, dv , \\
& & \int_{{\mathbb S}^2} (\nabla_v g) h \, dv = 2 \int_{{\mathbb S}^2} v g h \, dv -  \int_{{\mathbb S}^2} (\nabla_v h) g \, dv , \\
& & \int_{{\mathbb S}^2} (v \times \nabla_v g) h \, dv = -  \int_{{\mathbb S}^2} (v \times \nabla_v h) g \, dv ,
\end{eqnarray*}
for any pair of scalar functions $g$, $h$ on ${\mathbb S}^2$. We get: 
\begin{eqnarray*} 
{\mathcal L}^{(k)} &=& \frac{1}{c_1} \left( \int_{{\mathbb S}^2} \left[ - 2v F_\Omega + \nabla_v F_\Omega - \alpha (v \times \nabla_v F_\Omega) \right]  \psi^{(k)}\, dv  \right) \cdot \ell  .
\end{eqnarray*}
Now, we note that $\nabla_v F_\Omega = \beta (P_{v^\bot} \Omega)  F_\Omega$, which leads to 
\begin{eqnarray*} 
{\mathcal L}^{(k)} &=& \frac{1}{c_1} \left( \int_{{\mathbb S}^2} \left[ - 2v  + \beta (P_{v^\bot} \Omega) - \alpha \beta (v \times \Omega) \right]  \psi^{(k)} \, F_\Omega \, dv  \right) \cdot \ell  ,
\end{eqnarray*}
(we used that $v \times \Omega = v \times (P_{v^\bot} \Omega) $).
We use the decomposition $v=v_\bot + v_\parallel$. Since $\ell \bot \Omega$ and $P_{\Omega^\bot} P_{v^\bot} \Omega = - (v \cdot \Omega) v_\bot$, we can write: 
\begin{eqnarray*} 
{\mathcal L}^{(k)} &=& \frac{1}{c_1} \left( \int_{{\mathbb S}^2} \left[ - ( 2 + \beta (v \cdot \Omega)) v_\bot - \alpha \beta (v_\bot \times \Omega) \right]  \psi^{(k)} \, F_\Omega \, dv  \right) \cdot \ell  .
\end{eqnarray*}
Introducing the coordinates $\ell = (\ell_1, \ell_2, 0)$ in the cartesian basis associated to $\Omega$, and decomposing $v_\bot$ accordingly, we get: 
\begin{eqnarray*} 
{\mathcal L}^{(k)} &=& \frac{1}{c_1} \int_{{\mathbb S}^2} \left[  - ( 2 + \beta \cos \theta) (\cos \varphi \ell_1 + \sin \varphi \ell_2) \right. \\
& & \hspace{4cm} \left. - \alpha \beta  (\sin \varphi \ell_1 - \cos \varphi \ell_2) \right]  \psi^{(k)} \, F_\Omega \, \sin \theta \, dv    \\
 &=& \frac{1}{c_1}  \int_{{\mathbb S}^2} \left[  ( - ( 2 + \beta \cos \theta) \ell_1 + \alpha \beta \ell_2) \cos \varphi  + \right.  \\
& & \hspace{4cm} \left. + ( - ( 2 + \beta \cos \theta) \ell_2 - \alpha \beta \ell_1 ) \sin \varphi \right]  \psi^{(k)} \, F_\Omega \, \sin \theta \, dv .  
\end{eqnarray*}
Finally, using (\ref{eq_psik_1}), (\ref{eq_psik_2}), we get:
\begin{eqnarray*} 
{\mathcal L}^{(1)} &=& \frac{1}{c_1} \int_{{\mathbb S}^2}  \left[  ( - ( 2 + \beta \cos \theta) \ell_1 + \alpha \beta \ell_2) \cos \varphi  + ( - ( 2 + \beta \cos \theta) \ell_2 - \alpha \beta \ell_1 ) \sin \varphi \right] \\
& & \hspace{6cm}  (\psi_1(\theta) \cos \varphi + \psi_2(\theta) \sin \varphi) \, F_\Omega \, \sin \theta \, dv  ,  \\
{\mathcal L}^{(2)} &=& \frac{1}{c_1}  \int_{{\mathbb S}^2} \left[  ( - ( 2 + \beta \cos \theta) \ell_1 + \alpha \beta \ell_2) \cos \varphi  + ( - ( 2 + \beta \cos \theta) \ell_2 - \alpha \beta \ell_1 ) \sin \varphi \right] \\
& & \hspace{6cm} (- \psi_2(\theta) \cos \varphi + \psi_1(\theta) \sin \varphi) \, F_\Omega \, \sin \theta \, dv ,  
\end{eqnarray*}
and performing the integration with respect to $\varphi \in [0,2\pi]$, we are led to
\begin{eqnarray*} 
{\mathcal L}^{(1)} &=& \frac{1}{2c_1} \int_0^\pi \left[  ( - ( 2 + \beta \cos \theta) \ell_1 + \alpha \beta \ell_2) \psi_1(\theta)  + ( - ( 2 + \beta \cos \theta) \ell_2 - \alpha \beta \ell_1 ) \psi_2(\theta) \right] \\
& & \hspace{10cm} \, F_\Omega \, \sin^2 \theta \, d \theta  ,  \\
{\mathcal L}^{(2)} &=& \frac{1}{2c_1} \int_0^\pi \left[  - ( - ( 2 + \beta \cos \theta) \ell_1 + \alpha \beta \ell_2) \psi_2(\theta)  + ( - ( 2 + \beta \cos \theta) \ell_2 - \alpha \beta \ell_1 ) \psi_1(\theta) \right] \\
& & \hspace{10cm} \, F_\Omega \, \sin^2 \theta \, d \theta .  
\end{eqnarray*}
In vector notations, this is written: 
\begin{eqnarray*}
&& \hspace{-1cm} \left( \begin{array}{c} {\mathcal L}^{(1)} \\ {\mathcal L}^{(2)} \\ 0 \end{array} \right) = \frac{1}{c_1} \left\{ - (2 [A] + \beta [B]) \ell + \alpha \beta [A] (\ell \times \Omega) \right\} . 
\end{eqnarray*}

Finally, collecting all terms leads to the following equation: 
\begin{eqnarray*}
&& \hspace{-1cm} [A] P_{\Omega^\bot} ( c \nabla_x \rho + \beta \rho \partial_t \Omega) + \beta \rho c [B] P_{\Omega^\bot} ((\Omega \cdot \nabla_x) \Omega) + \\
&& \hspace{5cm}
+ \frac{1}{c_1} \left\{ - (2 [A] + \beta [B]) \ell + \alpha \beta [A] (\ell \times \Omega) \right\} =0 . 
\end{eqnarray*}
Now, multiplying the first two lines of this vector equation by $\beta [a]^{-1}$, we get
\begin{eqnarray*}
&& \hspace{-1cm} \rho (\partial_t \Omega + c [C_2] (\Omega \cdot \nabla_x) \Omega ) + c d P_{\Omega^\bot} \nabla_x \rho + \frac{1}{c_1} \left\{ - (2d \mbox{Id} + [C_2]) \ell + \alpha (\ell \times \Omega) \right\} = 0 , 
\end{eqnarray*}
or, expliciting the expression of $\ell$: 
\begin{eqnarray*}
&& \hspace{-1cm} \rho (\partial_t \Omega + c [C_2] (\Omega \cdot \nabla_x) \Omega ) + c d P_{\Omega^\bot} \nabla_x \rho + \frac{1}{c_1} \left\{ - (2d \mbox{Id} + [C_2]) P_{\Omega^\bot} ( k c_1 \Delta (\rho \Omega) - \phi \nabla_x \rho ) + \right. \\
&& \hspace{7cm} + \left. \alpha ( ( k c_1 \Delta (\rho \Omega) - \phi \nabla_x \rho ) \times \Omega) \right\} = 0 , 
\end{eqnarray*}
Using (\ref{eq_rotation}), we can also write this equation in the form (\ref{eqnl:omega_limit}), which ends the proof of the theorem. \endproof

\setcounter{equation}{0}
\section{Conclusion}
\label{sec:conclu}

In this paper, we have derived the hydrodynamic limit of a kinetic model of self-propelled particles with alignment interaction and with precession about the alignment direction. We have shown that the resulting system consists of a conservative equation for the local density and a non-conservative equation for the orientation. We have observed that this system may lose its hyperbolicity. Then, we have provided an extension of the model including diffusion terms under the assumption of weakly non-local interaction. In the particular case of zero self-propelling speed, we have noted that the resulting model is nothing but the classical Landau-Lifschitz-Gilbert equation. Therefore the present theory provides one of the very few (if not the only) kinetic justification of the phenomenological Landau-Lifschitz-Gilbert equation. Future works in this direction will include the sensitivity analysis of the hydrodynamic model in terms of the modeling parameters, the accounting for phase transitions, the influence of a vision angle, and the inclusion of the particle interactions with a surrounding fluid in the perspective of modeling active particles suspensions. Another research direction consists in investigating the mathematical structure of these models, prove well-posedness and investigate its qualitative properties.


%

\end{document}